\documentclass[10pt,twocolumn,osajnl2,showpacs,superscriptaddress, final]{revtex4}  
\usepackage{graphicx}
\usepackage[english]{babel}
\usepackage[pdftex,colorlinks]{hyperref} 

\DeclareGraphicsExtensions{.pdf,.jpg}

\begin{document}

\title{A simple integrated single atom detector}

\author{Marco Wilzbach}
\affiliation{Atominstitut der {\"O}sterreichischen Universit{\"a}ten, TU-Wien, Vienna, Austria}
\affiliation{Physikalisches Institut, Universit{\"a}t Heidelberg, Heidelberg, Germany}

\author{Dennis Heine}
\affiliation{Atominstitut der {\"O}sterreichischen Universit{\"a}ten, TU-Wien, Vienna, Austria}
\affiliation{Physikalisches Institut, Universit{\"a}t Heidelberg, Heidelberg, Germany}

\author{S{\"o}nke Groth}
\affiliation{Physikalisches Institut, Universit{\"a}t Heidelberg, Heidelberg, Germany}

\author{Xiyuan Liu}
\affiliation{Lehrstuhl f{\"u}r Optoelektronik, Universit{\"a}t Mannheim, Mannheim, Germany}

\author{Thomas Raub}
\affiliation{Atominstitut der {\"O}sterreichischen Universit{\"a}ten, TU-Wien, Vienna, Austria}
\affiliation{Physikalisches Institut, Universit{\"a}t Heidelberg, Heidelberg, Germany}

\author{Bj{\"o}rn Hessmo}
\email{hessmo@atomchip.org}
\affiliation{Atominstitut der {\"O}sterreichischen Universit{\"a}ten, TU-Wien, Vienna, Austria}
\affiliation{Physikalisches Institut, Universit{\"a}t Heidelberg, Heidelberg, Germany}

\author{J{\"o}rg Schmiedmayer}
\affiliation{Atominstitut der {\"O}sterreichischen Universit{\"a}ten, TU-Wien, Vienna, Austria}
\affiliation{Physikalisches Institut, Universit{\"a}t Heidelberg, Heidelberg, Germany}

\begin{abstract}%
We present a reliable and robust integrated fluorescence detector capable of detecting single atoms. The detector consists of a tapered lensed single-mode fiber for precise delivery of excitation light and a multimode fiber to collect the fluorescence. Both are mounted in lithographically defined SU-8 holding structures on an atom chip. ${}^{87}$Rb atoms propagating freely in a magnetic guide are detected with an efficiency of up to 66\% and a signal to noise ratio in excess of 100 is obtained for short integration times.
\end{abstract}

%

\maketitle

%
In an ideal background free fluorescence detector a single detected photon indicates the presence of atoms in the detection region. Moderate backgrounds can be tolerated if the atom remains localized, allowing to collect many fluorescence photons \cite{Grangier:SingleAtomEmission,Diedrich:AntibunchingfromIon,Blatt:IonTraps}. Detection of moving particles with finite interaction time with the detector requires supreme background suppression to reach high detection efficiencies \cite{bondo:2006}. 
Cavities can be used to enhance the fluorescence or absorption signal \cite{Esslinger:AtomLaser,AokiKimble:Microresonator,ColombeReichel:BECinFibreCavityonChip,Rempe:SingleAtomCavity,teper:CavityonChip,Haase:cavity,trupke:Cavity} but these detection schemes require active stabilization, increasing the complexity of the setup.

Our approach differs. We present here an integrated fluorescence detector which reaches high single atom detection efficiency without the need for either localization of the atoms or assistance of a cavity.

\begin{figure}[ht!]
	\centerline{\includegraphics{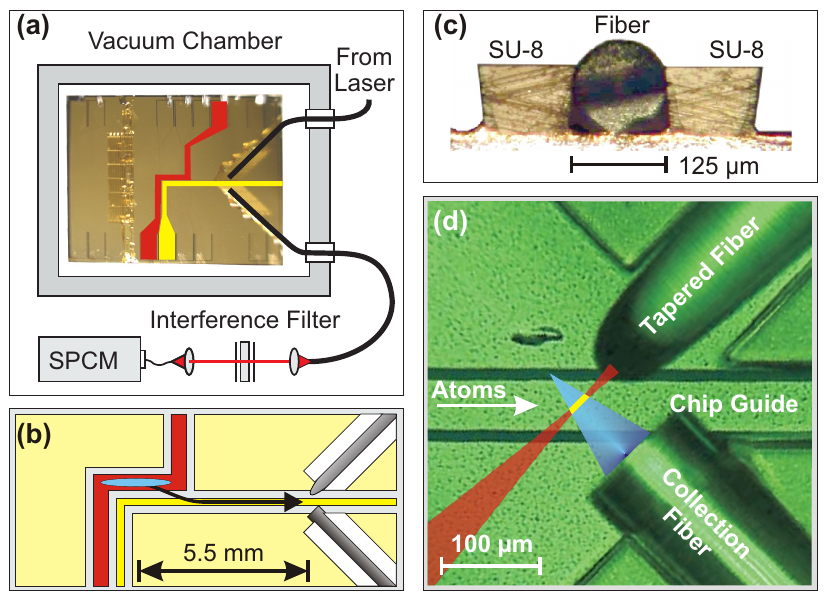}}
	\caption{\label{fig:setup}(color online) The integrated detector: 
	 (a) Basic layout of the detector and the atom chip. 
	 Resonant excitation light is delivered via a tapered lensed fiber.
    The fluorescence light is collected by a multimode fiber and
    guided to a single photon counting module (SPCM).
   (b) Schematic layout of the atom chip.
    Atoms are initially trapped in a magnetic trap generated by a Z-shaped wire (red). A magnetic guide (yellow) transports the atoms to the focus of the tapered lensed fiber.
   (c) Cross-section through the SU-8 structures holding the fiber.
   (d) Top view of the fibers. The multimode fiber collects fluorescence light from the excited atoms (blue cone).}
\end{figure}
%

%
The detector consist of a single-mode tapered lensed excitation fiber and a multimode detection fiber (Fig.\ref{fig:setup}(d)). The excitation fiber delivers resonant probe light to a focal spot of $5~\mu$m diameter $62.5~\mu$m above the wire defining the atom guide.
The collection fiber (NA 0.275) collects $1.9\%$ of the fluorescence photons and guides them via a feed-through \cite{Abraham:TeflonFeedthrough} out of the vacuum chamber.
To eliminate background light and to protect the photon counter, the light passes through an interference filter before being directed to the single photon counting module (SPCM) for a total photon detection efficiency of $0.9\%$ (calculated from known loss channels and detector efficiencies).
The fibers are arranged at $45^{\circ}$ to the guide and orthogonal to one another.
In the absence of atoms, less than $10^{-8}$ of the excitation light is scattered into the collection fiber, typically less than 10 photons per second in total. Due to the highly selective excitation and a matched collection region a total background of only $311~$counts per second (cps) is reached, dominated by the SPCM dark count rate of $245$~cps.

The detector is integrated on an atom chip \cite{groth:fabrication} by mounting the fibers in lithographically defined trenches.
These structures are fabricated from SU-8, an epoxy based
photoresist which allows the production of thick structures with very smooth sidewalls \cite{FortPhys:2006,liu2005}. The SU-8 fabrication is fully compatible with our atom chip production process \cite{groth:fabrication}. With a trench height of $90~\mu$m and slightly undercut walls the fibers ($125~\mu$m diameter) are clamped down onto the chip (Fig.\ref{fig:setup}(c)). The trenches allow accurate and stable passive alignment of the fibers with a precision of a few ten nanometres.
Temperature changes and gradients up to $100^{\circ}$C resulted in no measurable misalignment of the fibers. Long term stability under experimental conditions of more than one year has been observed.

%
\begin{figure}[tb]
	\centerline{\includegraphics{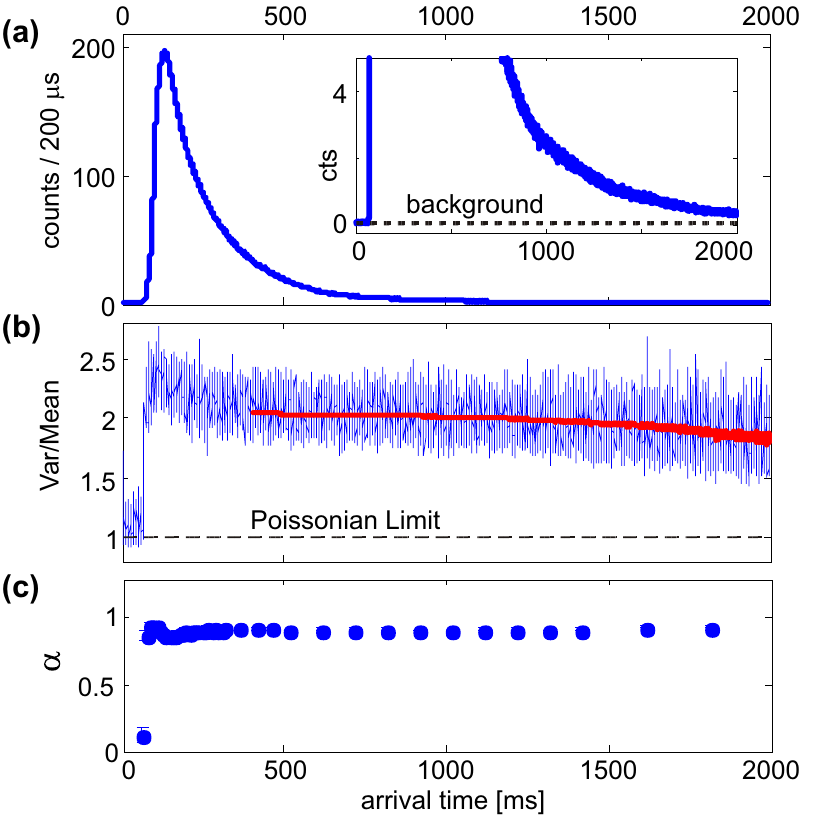}}
	\caption{\label{fig:signal}(color online) Photon statistics:
	(a) The mean photon count rate averaged over 600 individual measurements represents the density profile of the atom pulse passing the detector (integration time $200~\mu$s). The black dotted line indicates the mean background level. 
	(b) The ratio variance over mean shows that for the first 50~ms, where no atoms are present, the collected background follows Poissonian statistics. As soon as the atoms arrive $\mathrm{Var}(n)/\left\langle n \right\rangle$ exhibits a superpoissonian value of $1+\alpha$. Even though $\alpha$ remains constant the ratio decreases with atomic density as background becomes more important. The red line gives a fit according to Eq.\ref{eq:varmeanfull} over more than two orders of magnitude in atomic density. The initial overshoot is an artifact caused by the extreme slope around peak intensities due to the employed integration time.
	(c) Signal strength $\alpha$ as function of arrival time illustrating that $\alpha$ is independent of atom density (integration time $50~\mu$s).}
\end{figure}
In our experiment we first load neutral ${}^{87}$Rb atoms into a magneto-optic trap (MOT). Once the MOT is saturated with atoms the trap is shifted close to the atom chip and the atoms are transferred into a magnetic trap generated by a Z-shaped wire on the atom chip (see Fig.\ref{fig:setup}(a) and (b)) \cite{wildermuth:PRA,Zimmermann:2007}. 
From the trap the atoms are transferred into a magnetic guide generated by a L-shaped wire in which the atoms expand towards the detection region according to their temperature ($\approx 40\mu$K for the data presented here). The current through this wire is adjusted to overlap the minimum of the guide potential with the focal spot where the atoms are excited.

One striking observation from measurements with this detector is that stray light effects on the guided atoms can be neglected. This is quite remarkable, because magnetic traps and guides are extremely sensitive to the presence of light close to resonance with the atomic transition. Scattering of a little more than a single photon is sufficient to pump the atom into a magnetically untrapped state, expelling it from the magnetic guide.

The atoms passing the detector creates a fluorescence signal shown in Fig.\ref{fig:signal}(a) from which the atom density can be inferred if the number of counts per atom (cpa) $\alpha$ is known.
$\alpha$ can be determined from the photon statistics. A coherent light source of constant mean photon number $\left\langle n \right\rangle$ would create a Poisson distributed photon stream with variance equal to $\left\langle n \right\rangle$ and hence $\mathrm{Var}(n)/\left\langle n \right\rangle = 1$. Because the atom number fluctuates according to a distribution $p_{\mathrm{at}}(m)$ Mandel's formula \cite{RMP:MandelWolf} has to be used to calculate the ratio of variance to mean to
\begin{equation}
   \frac{\mathrm{Var}(n)}{\left\langle n \right\rangle} = 1 + \alpha{} \frac{\mathrm{Var}(m)}{\left\langle m \right\rangle + \left\langle bg \right\rangle / \alpha}
\label{eq:varmeanfull}
\end{equation}
where a Poissonian background noise source with mean intensity $\left\langle bg \right\rangle$ is included.
Fig.\ref{fig:signal}(b) shows the measured ratio $\mathrm{Var}(n) / \left\langle n \right\rangle$ for a thermal atomic ensemble which follows a Poissonian distribution $\mathrm{p}(m)$. In the absence of atoms only a Poissonian background is measured. As soon as the atoms arrive at the detector the noise increases to $\mathrm{Var}(n) / \left\langle n \right\rangle = 1 + \alpha$. From these measurements the signal strength can be determined to $\alpha=1.08$ cpa for an integration time of $t_\mathrm{int}=300~\mu$s, independent of atomic density (see Fig.\ref{fig:signal}(c)).

\begin{figure}[tb]
	\centerline{\includegraphics{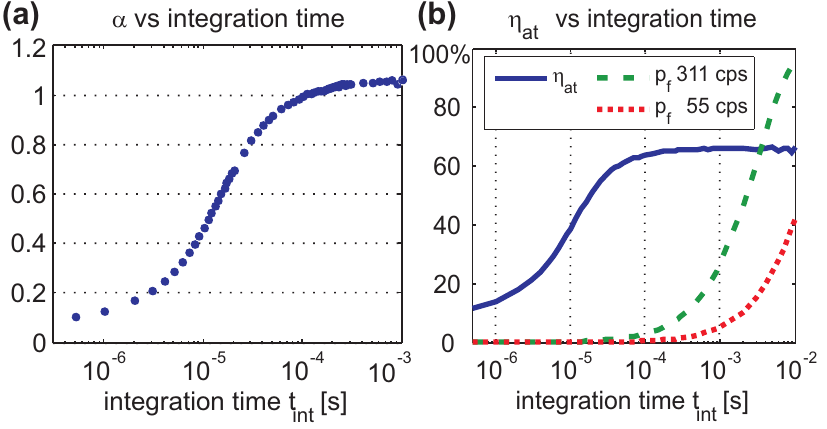}}
	\caption{\label{fig:efficiency}(color online) Effect of integration time:
	(a) Signal strength $\alpha$ as function of the integration time $t_\mathrm{int}$. While $300~\mu$s are needed to collect the full signal $\alpha$ drops only slowly for shorter integration times.
	(b) Single atoms detection efficiency $\eta_{\mathrm{at}}$ (solid line) and false positive detection probability $\mathrm{p}_{\mathrm{f}}$ for a background of 311 cps (dashed, current setup) and 55 cps (dotted, improved setup) shown versus $t_\mathrm{int}$.}
\end{figure}
The single atom detection efficiency is given by $\eta_{\mathrm{at}}=1-\exp{(-\alpha)}$, where $\exp{(-\alpha)}$ is the probability of detecting zero photons when an atom is present in the detection region. The detection efficiency  reaches 66\% for $300~\mu$s integration time. Background counts during this time lead to a false positive detection probability of $\mathrm{p}_\mathrm{f}=9\%$. 
Fig.\ref{fig:efficiency}(a) shows the signal strength $\alpha$ as a function of the integration time, $t_\mathrm{int}$. At the same time, the mean number of background counts $\left\langle bg \right\rangle$ is a linear function of $t_\mathrm{int}$. For most applications it is therefore desirable to work at shorter integration times leading to high signal to noise ratios (SNR) (see Fig.\ref{fig:efficiency}(b)). Defining $\mathrm{SNR}=\alpha(t_\mathrm{int}) / \left\langle bg(t_\mathrm{int}) \right\rangle$ an $\mathrm{SNR}=100$ is reached at $t_\mathrm{int}=25~\mu$s ($\eta_{\mathrm{at}}=55\%$). At $t_\mathrm{int}=45~\mu$s the atom detection efficiency is $\eta_{\mathrm{at}}=60\%$ with only $\mathrm{p}_{\mathrm{f}}=1.4\%$ false detection probability (Tab.\ref{tab:detefficiency}). The capability of detecting single atoms was independently verified by observing near-perfect anti-bunching of the photons with $g^{(2)}(0)\sim 0.05$

The total number of photons scattered by the atoms before leaving the detection region can be independently obtained by measuring the ratio of fluorescence counts generated by illumination on the F=2$\rightarrow$F'=1 and the F=2$\rightarrow$F'=3 transitions. On the F=2$\rightarrow$F'=1 transition an atom scatters slightly more than one photon before being optically pumped into the other hyperfine ground state.
This ratio indicates that each atom scatters $\sim{}120$ photons on the employed cyclic F=2$\rightarrow$F'=3 transition before it leaves the detector. This confirms the above given photon detection efficiency of \mbox{$\eta_\mathrm{ph}=0.9$\%}. In addition the value of $\alpha$ was confirmed by independent global atom number
measurements using standard absorption imaging.

%
The efficiency of the detector is limited by the NA of the collection fiber, while the SNR is limited by the background. Exchanging the employed photon detector (Perkin-Elmer SPCM-AQR-12-FC) by a low noise model with a dark count rate $<25$ cps a total background of $55$ cps can be achieved. Here $\mathrm{SNR}=100$ would be reached already for $t_\mathrm{int}=175~\mu$s with $\eta_{\mathrm{at}}=65\%$ and $\mathrm{p}_{\mathrm{f}}=1\%$.
Exchanging the collection fiber by a commercially available fiber of $\mathrm{NA}=0.53$ up to $\alpha=4.5$ is expected and $\eta_{\mathrm{at}}\approx{}95\%$ could be reached for $t_\mathrm{int}=20~\mu$s. Two detection fibres with $\mathrm{NA}=0.53$ a single atom detection efficiency beyond $99\%$ for $t_\mathrm{int}=20~\mu$s seems feasible. Such a system would reach $\alpha=9$. This allows true atom counting by transient count rate analysis \cite{bondo:2006}.

%
\begin{table}[tb]
	\centering
		\begin{tabular}{|l|ccc|}
			\multicolumn{4}{c}{single atom detection efficiency (NA 0.275 (0.53))}\\
			\hline
			$t_{\mathrm{int}}$			& $300~\mu$s	& $45~\mu$s	& $20~\mu$s\\
			$\alpha$ [cpa]					& $1.08$ $(4.5)$	& $0.92$ $(3.8)$& $0.72$ $(3.0)$\\
			$\eta_{\mathrm{at}}$		& $66\%$ $(99\%)$	& $60\%$ $(98\%)$& $50\%$ $(95\%)$\\
			$\mathrm{p}_{\mathrm{f}}$& $9\%$ $(<4.3\%)$	& $1.4\%$ $(<0.7\%)$& $0.6\%$ $(<0.3\%)$\\
			\hline
		\end{tabular}
		\caption{Single atom detection efficiency as function of the integration time for the current system, $\eta_{\mathrm{ph}}=0.9\%$. Numbers in brackets denote the projected values for an improved system with a fiber of NA 0.53 \mbox{$\eta_{\mathrm{ph}}=3.8\%$ and low noise photon detector.}%
		\label{tab:detefficiency}}
\end{table}
%

%
We thank A.~Haase, and M.~Schwarz for help in the early stages of the experiment, and K.-H.~Brenner, and I.~Bar-Joseph for support in fabrication. We gratefully acknowledge financial support from the Landesstiftung Baden-W{\"u}rttemberg, the European Union (SCALA), and the Austrian Nano Initiative (PLATON).


\begin{thebibliography}{10}
\newcommand{\enquote}[1]{``#1''}

\bibitem{Grangier:SingleAtomEmission}
B.~Darquie, M.~Jones, J.~Dingjan, J.~Beugnon, S.~Bergamini, Y.~Sortais,
  G.~Messin, A.~Browaeys, and P.~Grangier, Science \textbf{309}, 454 (2005).

\bibitem{Diedrich:AntibunchingfromIon}
F.~Diedrich and H.~Walther, Phys. Rev. Lett. \textbf{58}, 203 (1987).

\bibitem{Blatt:IonTraps}
D.~Leibfried, R.~Blatt, C.~Monroe, and D.~Wineland, Rev. Mod. Phys.
  \textbf{75}, 281 (2003).

\bibitem{bondo:2006}
T.~Bondo, M.~Hennrich, T.~Legero, G.~Rempe, and A.~Kuhn, Opt. Commun.
  \textbf{$264$}, $271$ ($2006$).

\bibitem{Esslinger:AtomLaser}
A.~{\"O}ttl, S.~Ritter, M.~K{\"o}hl, and T.~Esslinger, Phys. Rev. Lett.
  \textbf{95}, 090404 (2005).

\bibitem{AokiKimble:Microresonator}
T.~Aoki, B.~Dayan, E.~Wilcut, W.~Bowen, A.~Parkins, T.~Kippenberg, K.~Valhala,
  and H.~Kimble, Nature \textbf{443}, 671 (2006).

\bibitem{ColombeReichel:BECinFibreCavityonChip}
Y.~Colombe, T.~Steinmetz, G.~Dubois, F.~Linke, D.~Hunger, and J.~Reichel,
  Nature \textbf{450}, 272 (2007).

\bibitem{Rempe:SingleAtomCavity}
P.~M\"unstermann \emph{et~al.}, Phys. Rev. Lett. \textbf{82}, 3791 (1999).

\bibitem{teper:CavityonChip}
I.~Teper, Y.-J. Lin, and V.~Vuleti\'{c}, Phys. Rev. Lett. \textbf{97}, 023002
  (2006).

\bibitem{Haase:cavity}
A.~Haase, B.~Hessmo, and J.~Schmiedmayer, Opt. Lett. \textbf{31}, 268 (2006).

\bibitem{trupke:Cavity}
M.~Trupke, J.~Goldwin, B.~Darqui\'{e}, G.~Dutier, S.~Eriksson, J.~Ashmore, and
  E.~A. Hinds, Phys. Rev. Lett. \textbf{99}, 063601 (2007).

\bibitem{Abraham:TeflonFeedthrough}
E.~R. Abraham and E.~A. Cornell, Appl. Opt. \textbf{37}, 1762 (1998).

\bibitem{groth:fabrication}
S.~Groth, P.~Kruger, S.~Wildermuth, R.~Folman, T.~Fernholz, J.~Schmiedmayer,
  D.~Mahalu, and I.~Bar-Joseph, Appl. Phys. Lett. \textbf{85}, 2980 (2004).

\bibitem{FortPhys:2006}
M.~Wilzbach, A.~Haase, M.~Schwarz, D.~Heine, K.~Wicker, S.~G. X.~Liu,
  K.-H.~Brenner, T.~Fernholz, B.~Hessmo, and J.~Schmiedmayer, Fortschr. Phys.
  \textbf{$54$}, $746$ ($2006$).

\bibitem{liu2005}
X.~Liu, K.-H. Brenner, M.~Wilzbach, M.~Schwarz, T.~Fernholz, and
  J.~Schmiedmayer, Appl. Opt. \textbf{44}, 6857 (2005).

\bibitem{wildermuth:PRA}
S.~Wildermuth, P.~Kr\"uger, C.~Becker, M.~Brajdic, S.~Haupt, A.~Kasper,
  R.~Folman, and J.~Schmiedmayer, Phys. Rev. A \textbf{69}, 030901 (2004).

\bibitem{Zimmermann:2007}
J.~Fort\'{a}gh and C.~Zimmermann, Rev. Mod. Phys. \textbf{79}, 235 (2007).

\bibitem{RMP:MandelWolf}
L.~Mandel and E.~Wolf, Rev. Mod. Phys. \textbf{37}, 231 (1965). To calculate
  the mean and variance we use
  \protect{\mbox{$p_{\mathrm{photon}}(n)=\sum_{m=0}^{\infty}p_{\mathrm{at}}(m)%
\frac{(\alpha m)^n}{n!}e^{-\alpha m}$}}, where $p_{\mathrm{at}(m)}$ is the
  probability to have $m$ atoms in the considered time interval. Each atom
  emits on average $\alpha$ photons.

\end{thebibliography}

\end{document}